\documentclass[aps,twocolumn,amsmath,amsfonts,nofootinbib,preprintnumbers,
  superscriptaddress,secnumarabic]{revtex4-1}
\usepackage[utf8]{inputenc}
\usepackage{color,graphicx,hyperref,amsmath,amssymb,multirow,slashed}
\hypersetup{colorlinks=true,linkcolor=blue,citecolor=magenta,filecolor=magenta,
  urlcolor=cyan}
\usepackage{enumitem}

\begin{document}
\title{Probing heavy scalars with an effective $Hb\bar bg$ coupling at the LHC}
\author{Katri Huitu}
\email{ katri.huitu@helsinki.fi}
\affiliation{Department of Physics, and Helsinki Institute of Physics, P. O. Box 64, FI-00014 University of Helsinki, Finland}
\author{Subhadeep Mondal}
\email{subhadeep.mondal@helsinki.fi}
\affiliation{Department of Physics, and Helsinki Institute of Physics, P. O. Box 64, FI-00014 University of Helsinki, Finland}
\author{Biswarup Mukhopadhyaya}
\email{biswarup@iiserkol.ac.in }
\affiliation{Department of Physical Sciences, Indian Institute of Science Education and Research, Kolkata, Mohanpur 741246, 
Nadia, West Bengal, India}
\begin{abstract}
We have explored the prospect of probing a neutral scalar ($H$) produced in association with one $b$-quark and decaying either 
invisibly or into a pair of $b$-quarks at the LHC with centre of mass energy $\sqrt s = 14$ TeV. In this regard, we 
adopt an effective theory approach to parameterize a $Hb\bar bg$ vertex arising from a dimension six operator that 
encompasses the effect of some new physics setting in at a high scale. We concentrate solely on the five-flavor scheme 
to ascertain the sensitivity of the 14 TeV LHC in probing such an effective coupling as a function of the scalar mass at 
the highest possible projected luminosity, $3000~{\rm fb}^{-1}$. Through our multivariate analysis using machine learning 
algorithm we show that staying within the perturbative limit of the Wilson coefficient of the effective interaction, 
evidence with statistical significance of $3\sigma$ can be obtained in two different signal regions 
for $m_H\lesssim 2$ TeV and the scale of new physics $\Lambda = 3$ TeV. 
\end{abstract}
\maketitle
\section{Introduction}
\label{sec:intro}
In the quest for new physics beyond the standard model (BSM) heavy scalar search holds a particular importance. A bunch of new physics
models predict an extended scalar sector compared to the SM. In the absence of any direct hint of such non-standard scalar particles, 
one can only exclude certain regions of parameter space from the experimental data \cite{Aaboud:2018knk,Sirunyan:2018zkk,Sirunyan:2018qca,Sirunyan:2018iwt,Aaboud:2018ftw,Aaboud:2018sfw,Aad:2019yxi,Aaboud:2018sfi}. 
Indirect bounds on these particle masses and their couplings can also be derived from precision electroweak data, 
see e.g. \cite{Chowdhury:2017aav}. Under these circumstances, lack of clarity regarding specific new 
physics models has aroused considerable interest in effective theories \cite{Buchmuller:1985jz,Grzadkowski:2010es,Dedes:2017zog}. 
One can parameterize effects of new physics theories through masses and couplings of the particle in consideration. The results 
derived this way are model independent, and applicable to any new physics theory that can give rise to similar phenomenology. 

The $b$-quark distribution in the proton is non-negligible at the Large Hadron Collider (LHC). However, its role in the production
of the 125 GeV scalar is not estimated to be very significant.
On the other hand, with the ever-alive interest in an extended electroweak symmetry breaking sector, it is important 
to see if any heavier, additional neutral spinless particle can have enhanced production rate due to its $b$-coupling. 
In two Higgs doublet models (2HDM), for example, this is possible for appropriate values of $\tan\beta$, the ratio of the two doublet 
vacuum expectation values (vev) and $\alpha$, the doublet mixing angle. One may integrate this idea into a model-independent approach, 
and see if enhanced production rates follow in terms of some suitably parameterized interaction with a $b$-quark pair. This may yield 
new search channels for additional scalars, and also lead to constraints of the corresponding interactions and scalar masses.

In this paper we look at a non-standard scalar particle ($H$) and its effective coupling to a pair of bottom quarks and a gluon generated 
through a dimension six effective operator \cite{Grzadkowski:2010es}. The considered vertex can give rise to a possible production 
mode for the heavy scalar with non-negligible cross-section. 
Experimental collaborations have studied the production of a heavy scalar associated with bottom quarks 
to look for direct evidence of the scalar decaying into $b\bar b$ resulting in a multi $b$-jet final state 
\cite{Sirunyan:2018taj,Aad:2019zwb}. Such a computation requires careful inspection of both the four flavor 
and five flavor schemes \cite{Harlander:2011aa}. In the conventional four flavor scheme, the relevant production 
mode $pp\to b\bar b H$ is either quark or gluon initiated, whereas, $pp\to b H$ can also contribute 
to the same final state. The second production mode can only be obtained in a five-flavor scheme, since 
it is a bottom quark-gluon initiated process. 

Here we only concentrate on the five flavor scheme through a quark-quark-scalar-gluon effective 
vertex in order to ascertain how significant this contribution can be in probing a heavy scalar. 
The aforementioned experimental studies \cite{Sirunyan:2018taj,Aad:2019zwb} rely heavily on a large 
$Hb\bar b$ coupling. In the context of two Higgs doublet models this is a possibility in the alignment limit 
of ${\rm cos}(\beta - \alpha)\simeq 0$ \cite{Branco:2011iw}. In this limit, decay of the 
heavy scalar into the gauge bosons are suppressed and thus the LHC search of such a scalar heavily rely on its 
coupling to the third-generation fermions. In addition, if the $Hb\bar b$ coupling also is too small due to the presence of 
some new physics one would naturally expect the event rate into multi $b$-jet final 
state to go down. We, therefore, explore a scenario, where the cross-section $bg\to b H$ is mostly driven by 
the effective interaction which does not necessarily have to be small even if the $Hb\bar b$ coupling is 
small. Thus, in addition to probing the new effective interaction, one can also expect to improve the sensitivity  
of the $Hb\bar b$ coupling measurement. Once produced, we assume that $H$ can only decay through two 
possible decay modes, namely $b\bar b$ and $\chi\chi$, where $\chi$ represents some invisible particle like dark matter. 
The reason for concentrating only on these channels is that, in a sense, they correspond to the `difficult situations' 
in (heavy) Higgs identification at the two extremes. While a proliferation of $b$-jets in the final state threatens the event rate
due to suppression via identification efficiencies, one or two $b$'s with missing transverse energy (${\slashed E}_T$)
tends to get swamped by backgrounds. Thus, $H$ decaying invisibly leaves us with few options to probe. 
Vector boson fusion production channel cross-section can be small, for example in the alignment limit in 2HDM, which 
requires one to explore new possibilities. On the other hand, $H\to b\bar b$ is anticipated to be one of the most prominent 
decay modes of $H$ and also the corresponding coupling has significant impact on the production cross-section of $H$ in 
five flavor scheme.

In the four flavor scheme, the bottom 
quark mass ($m_b$) dependence is retained, while in five flavor scheme the b-quark is treated as a massless parton. 
Hence in the four flavor scheme, the bottom quark cannot appear as an initial state parton and can only arise from 
gluon splitting. The non-zero mass protects the gluon splittings from colinear divergences. In the five flavor scheme 
the computation of higher order correction to the strong coupling constant is simpler, since all the quarks are 
considered massless. Efforts have been going on in theoretical frontier to match these two schemes \cite{Forte:2015hba,
Bonvini:2015pxa,Bonvini:2016fgf,Forte:2016sja,Duhr:2020kzd}. The cross-section 
computed in the four flavor and five flavor schemes are more accurate in the limit $\frac{\mu}{m_b}\sim 1$ and $\frac{\mu}{m_b} \gg  1$, 
respectively, where $\mu$ represents the factorization and renormalisation scale \cite{Forte:2015hba,Forte:2016sja,Cacciapaglia:2018qep}. 
Since in this work we take the new physics scale to be 3 TeV and beyond, we stick to the five flavor scheme only.   
A novel aspect of the production mode explored here, $bg\to b H$, is that the $H$ is expected to have a large transverse momentum 
\cite{Arnesen:2008fb}. Such a scenario can be easily identified as a possible new 
physics signal. 

Over the last decade or so, the LHC analyses have slowly but surely been moving towards the use of machine learning methods to 
collect and analyze the large data samples \cite{Bhat:2010zz}. Algorithms like gradient boosted decision trees (BDTs) \cite{bdt:book1,bdt:book2} 
were used prior to the LHC experiment as well, see e.g. \cite{Roe:2004na}. Machine learning algorithms were highly useful in the discovery 
of the 125 GeV Higgs boson \cite{Aad:2012tfa,Chatrchyan:2012ufa} and subsequently with the accumulation of more and more data - analyzing them 
requires the use of machine learning algorithms more than ever. The BDT algorithms have widespread application in 
data analysis and can be used effectively in the context of collider physics to identify new physics signal events depending on the choices 
of suitable kinematic variables \cite{Baldi:2014kfa}. In this work we have adopted this technique and used the XGBoost toolkit \cite{Chen:2016btl}.

The paper is organized in the following way. In section~\ref{sec:obj} we briefly discuss the theoretical premise of our analysis 
and the impact of the existing experimental constraints. In section~\ref{sec:coll} we introduce the two different signal regions that we have 
studied in the context of LHC. We discuss in some detail the choice of our kinematic variables for the multivariate analyses. We also highlight 
the advantage of such analyses over the canonical cut-based approach. In section~\ref{sec:res_mult} we discuss the impact of our analyses via 
some chosen benchmark points and subsequently in section~\ref{sec:disc} we predict the highest achievable sensitivity of the 14 TeV LHC in probing 
the present scenario. Finally, in section~\ref{sec:concl} we summarize our work and draw conclusions.

\section{Theory framework and constraints}
\label{sec:obj}
As already mentioned, in this work we maintain a model independent approach. We parameterize the $Hb\bar bg$ vertex in terms 
of Wilson coefficient that encapsulates the effect of high scale theory. Here $H$ represents a heavy BSM scalar. The interaction term 
follows from dimension six $SU(2)\times U(1)$ gauge invariant operator consistent with the assumption that its origin lies above the 
electroweak symmetry breaking scale. The effective interaction relevant to our study is 
$\frac{f}{\Lambda^2} \bar b\sigma^{\mu\nu} b \frac{\lambda_a}{2}G^a_{\mu\nu}H$,
where $f$ is the Wilson coefficient, $\Lambda$ represents the cut-off scale at which the new physics sets in, 
$\lambda_a$ represents the Gell-Mann matrices, $\frac{\lambda_a}{2}$ being the $SU(3)$ generators, and $G_{\mu\nu}$ is the $SU(3)$ field 
strength tensor. Such an interaction can follow from a dimension six operator of the form \cite{Grzadkowski:2010es}: 
\begin{eqnarray}
 {\mathcal O} = \frac{1}{\Lambda^2} (\bar q\sigma^{\mu\nu}d)\frac{\lambda_a}{2}G^a_{\mu\nu}\Phi + {\rm h.c.}  \nonumber
\end{eqnarray}
Here $\Phi$, $q$ and $d$ represent the new scalar, left handed quark doublet and right handed quark, respectively. 

\begin{figure}[h]
\includegraphics[width=8cm,height=3cm]{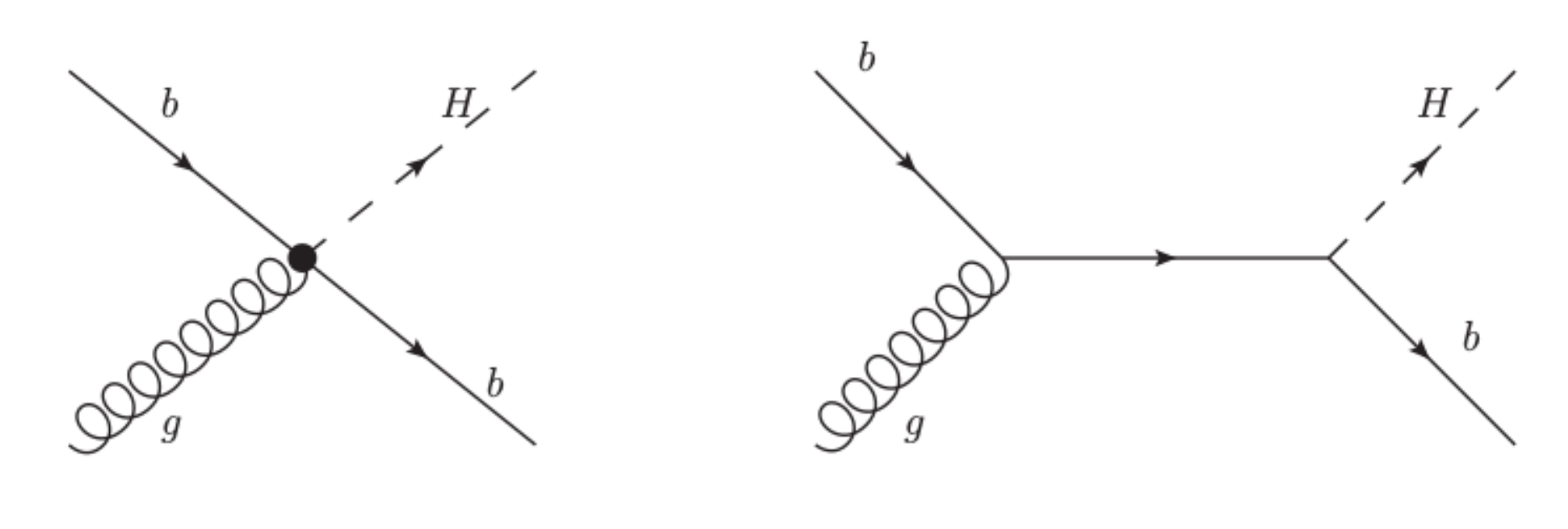}
\caption{Possible production modes for $bg\to bH$ in presence of the dimension six operator and a $Hb\bar b$ coupling.}
\label{fig:proc}
\end{figure}

Our objective is to probe the process $bg\to bH$ in the context of LHC in the five flavor scheme. A $Hb\bar b$ vertex 
can lead to similar phenomenology. The two possible production channels are shown in Fig.~\ref{fig:proc}.
We explore the possibility of $H$ either decaying into a pair of $b$-quarks or completely invisibly, e.g. into a pair of dark matter 
particles ($\chi$). Maintaining a model independent 
approach, we do not assume any specific coupling strengths for these two interactions. Instead we take the branching ratio, 
BR($H\to \chi\chi$), as a free parameter, which automatically fixes BR($H\to b\bar b$), since we assume there are no additional decay modes. 
The other relevant free parameters for our study are $f$, $\Lambda$, $m_H$ and $m_{\chi}$. The $H\to b\bar b$ coupling strength 
determines the contribution of the s-channel mediated diagram in Fig.~\ref{fig:proc} to the total cross-section and thus is also 
a relevant parameter when this cross-section becomes comparable to or dominant over the effective interaction diagram. 

Phenomenological constraints on the mass and couplings of such a scalar can be derived from the existing
direct search experimental analyses by the ATLAS and CMS Collaborations. However, they are subjected to 
its production cross-section and existing decay modes. 
Search for a BSM scalar usually involves the heavy scalar decaying into a pair of SM Higgs which eventually decay 
into various final states, like four b-jets \cite{Aaboud:2018knk, Sirunyan:2018zkk, Sirunyan:2018qca}, 
two b-jets along with photons \cite{Sirunyan:2018iwt, Aaboud:2018ftw}, pair of taus \cite{Aaboud:2018sfw} ,
or charged leptons and missing energy \cite{Aad:2019yxi}. The exclusion limit on the masses of such 
a scalar, therefore, depends upon its single production cross-section, e.g. through gluon fusion, 
and its coupling to a pair of SM Higgs bosons. These constraints are, thus, not directly applicable to the present scenario.
Production of scalars via bottom quark fusion has been studied in \cite{Sirunyan:2018taj,Aad:2019zwb}. These constraints are again subjected 
to the $Hb\bar b$ coupling strength. In this work, we assume this coupling to be orders of magnitude 
smaller compared to what is considered in these experimental studies, which ensures that we are well below 
the exclusion line throughout the range of $m_H$.

The most effective way to a direct search of invisible decay of a scalar is through a vector boson fusion production channel. 
A BSM scalar can be probed up to 3 TeV with the analyzed data so far \cite{Aaboud:2018sfi}. However, this involves the scalar 
coupling to a pair of gauge bosons. In our present scenario, for simplicity, we assume this coupling to be very weak 
and as a result such constraints are irrelevant.    
\section{Collider study: a multivariate analysis}
\label{sec:coll}
In this section, we study the collider physics implication of the present scenario. 
As already discussed, we focus on two possible decay modes of the heavy scalar, namely, $H\to\chi\chi$ and $H\to b\bar b$. 
For simplicity, we assume that these are the only two possible decay modes of $H$. As a result, the production channel 
$bg\to bH$ may lead to two possible signal regions, 1 b-jet + $\slashed{E}_T$ (SR1) and 3 b-jets (SR2). We explore 
both these final states in the context of 14 TeV LHC. For this, we perform a multivariate analysis  with machine learning 
technique known as the BDT. For our analysis, we have taken 1000 estimators, maximum depth 4 and learning rate 0.01. 
We combine data corresponding to 
chosen kinematic variables for our signal events as well as all the background events from different channels to prepare one 
single data file. The background events are combined after properly weighting them according to their cross-sections and it has 
been ensured that there are enough signal events to match the total weight of all the background events. 
We take $80\%$ of our combined data for training and 20$\%$ for testing purpose. 

We have used Madgraph5 \cite{Alwall:2011uj,Alwall:2014hca} for parton level event generation and 
PYTHIA8 \cite{Sjostrand:2006za,Sjostrand:2014zea} for subsequent showering and hadronisation. 
For the simulation, we use the NNPDF parton distribution function \cite{Ball:2012cx, Ball:2014uwa}.
We have used a dynamic factorisation and renormalisation scale. For a single heavy particle production it is 
simply $m^2 + p_T^2$, where, $m$ and $p_T$ represent the mass and transverse momentum of the particle.
For production of a pair of heavy particles it is the geometric mean of $m^2 + p_T^2$ for each particle \cite{scale}.
Detector simulation was performed using Delphes3 \cite{deFavereau:2013fsa,Selvaggi:2014mya,Mertens:2015kba}. 
Jets were constructed in FastJet \cite{Cacciari:2011ma} following anti-kt algorithm \cite{Cacciari:2008gp}. 
Appropriate $b$-jet tagging and mistagging efficiencies have been implemented within the Delphes platform following recommended 
values by the ATLAS Collabroration \cite{ATL-PHYS-PUB-2015-022}. 
Throughout this study, we have used next-to-leading order (NLO) cross-sections for all the SM processes \cite{Alwall:2014hca}. 
For the signal we assume that any new physics effect including the NLO contribution is clubbed with the Wilson coefficient of the dimension 
six operator.
\subsection*{Signal Region 1: 1 b-jet + $\slashed{E}_T$}
In this signal region $H$ decays invisibly, resulting in the sole $b$-jet associated with a large transverse missing energy. 
We put the following criteria to select events in this signal region: the final state should have only one $b$-jet with $p_T^b > 120$ GeV 
and $|\eta^j| < 2.5$, at least two light jets with $p_T^j > 40$ GeV and $|\eta^j| < 2.5$, no charged leptons with $p_T^{\ell} > 10$ GeV 
and $|\eta^{\ell}| < 2.5$, $\slashed{E}_T > 120$ GeV. We have studied the following SM background channels to estimate the efficiency of 
our analysis in this signal region: $t\bar t$+jets, $t$+jets, $V$+jets, $h_{\rm SM}$+jets and QCD jets, where $V$ represents the W and 
Z bosons.

The large $\slashed{E}_T$ is expected to be balanced by the sole $b$-jet $p_T$. We can use this property of the signal events to isolate 
them from the background. There can be additional radiation jets in the signal but they are likely to have much smaller transverse momenta, 
while most of the background channels appear with additional jets produced from decays of heavier particles like the top quark or a gauge 
boson. These jets are, therefore, expected to have harder $p_T$ distribution compared to the light jets in our signal. This can be another 
discriminatory factor. Top quark production channels pose a particular challenge to this signal region due to the large cross-section 
associated with them. However, the $\slashed{E}_T$ distribution is likely to differ from the signal subjected to the choice of $m_H$ and 
$m_{\chi}$. Moreover, the angular separation between the jet transverse momenta and the $\slashed{\vec{E}}_T$ in the signal are expected 
to be larger compared to that in the top quark production channels given the difference in decay topologies. Event shape analysis variables 
like transverse sphericity \cite{Banfi:2004nk,Abelev:2012sk} may also prove to be useful in this regard. The angular separation variables 
are also effective in 
reducing the potentially large contribution to the background originating from QCD multi-jet production modes. All these considerations 
lead us to choose the following kinematic variables to be used in our multivariate analysis. 
\begin{enumerate}[noitemsep]
 \item Number of light jets, $N_j$.
 \item Transverse momentum of the b-jet, $p_T^b$. 
 \item Transverse momentum of the hardest light jet, $p_T^{j1}$.
 \item Transverse momentum of the second hardest light jet, $p_T^{j2}$.
 \item Missing transverse energy, $\slashed{E}_T$.
 \item Angular separation between $\slashed{E}_T$ and hardest light jet, $\Delta\phi({\slashed{\vec{E}}_T}, \vec{p}_T^{j1})$. 
 \item Angular separation between $\slashed{E}_T$ and second hardest light jet, $\Delta\phi({\slashed{\vec{E}}_T}, \vec{p}_T^{j2})$.
 \item Angular separation between $\slashed{E}_T$ and the lone b-jet, $\Delta\phi({\slashed{\vec{E}}_T}, \vec{p}_T^b)$.
 \item Separation between the b-jet and hardest light jet, $\Delta R(b, j1)$.
 \item Separation between the b-jet and second hardest light jet, $\Delta R(b, j2)$.
 \item Effective mass, $M_{EFF}$.
 \item Transverse sphericity, $S_T = \frac{2\lambda_2}{\lambda_1 + \lambda_2}$, where, $\lambda_1$ and $\lambda_2$ are the eigenvalues of the matrix  
\begin{eqnarray}
\frac{1}{\sum_i p_{T_i}}\sum_i\begin{pmatrix} p^2_{x_i} & p_{x_i} p_{y_i} \\ p_{y_i} p_{x_i} & p^2_{y_i} \end{pmatrix} .\nonumber 
\end{eqnarray}
\end{enumerate}
\subsection*{Signal Region 2: 3 b-jet}
$H$ may also decay into a pair of $b$ quark resulting in a 3 $b$-jet final state with no direct source of missing energy. 
We put the following criteria to select events in this signal region: the final state should have three $b$-jet with 
a transverse momentum threshold of 25 GeV, while the hardest two $b$-jets must have $p_T^b > 80$ GeV, no charged leptons 
be used in our multivariate analysis.with $p_T^{\ell} > 10$ GeV. The $|\eta|$ criteria remain same as in SR1. The dominant SM backgrounds studied for this 
signal region are: $t\bar t$+jets, $t$+jets, $V$+jets, $VV$+jets, $bh_{\rm SM}$, $Zh_{\rm SM}$ and QCD jets.

Lack of any direct source of missing energy in this signal region is a disadvantage agaist the background channels like 
$V$ + jets and QCD multi-jet production channels. On the other hand, it can be a discriminating factor against background 
channels with large $\slashed{E}_T$, like $t\bar t$+jets. Among the three $b$-jets in the final state, a pair is generated 
from $H$ decay. This can be used to reconstruct $m_H$ to certain degree of accuracy and a window of invariant mass of this 
$b$-jet pair can be used as a discriminating variable. This $b$-jet pair is also likely to be well separated from the third 
$b$-jet. We choose the following list of kinematic variables to be used in our multivariate analysis to exploit these features.
\begin{enumerate}[noitemsep]
 \item Number of b-jets, $N_b$.
 \item Number of light jets, $N_j$.
 \item Transverse momentum of the hardest b-jet, $p_T^{b1}$.
 \item Transverse momentum of the second hardest b-jet, $p_T^{b2}$.
 \item Transverse momentum of the third hardest b-jet, $p_T^{b3}$.
 \item Missing transverse energy, $\slashed{E}_T$.
 \item Angular separation between $\slashed{E}_T$ and hardest  b-jet, $\Delta\phi({\slashed{\vec{E}}_T}, \vec{p}_T^{b1})$. 
 \item Angular separation between $\slashed{E}_T$ and second hardest  b-jet, $\Delta\phi({\slashed{\vec{E}}_T}, \vec{p}_T^{b2})$.
 \item Angular separation between $\slashed{E}_T$ and third hardest  b-jet, $\Delta\phi({\slashed{\vec{E}}_T}, \vec{p}_T^{b3})$.
 \item Separation between two hardest b-jets, $\Delta R(b1, b2)$.
 \item Invariant mass of hardest two b-jets, $m_{\rm inv}^{bb}$.
 \item Angular separation between the two hardest b-jet pair system and the third b-jet, $\Delta\phi(\vec{p}_T^{bb}, \vec{p}_T^{b3})$.
 \item Sum of against momentum of hardest three b-jets, $H^b_T$.
 \item Transverse sphericity, $S_T$.
\end{enumerate}
\subsection{Results}
\label{sec:res_mult}
For 1 b-jet + $\slashed{E}_T$ final state, the most important kinematic variables are $\slashed{E}_T$, $\Delta R(b, j1)$, $\Delta R(b, j2)$, 
$\Delta\phi({\slashed{\vec{E}}_T}, \vec{p}_T^b)$, $p_T^b$ and $M_{EFF}$. The $p_T$ of the light jets are also essential in obtaining the 
resultant sensitivity. $N_j$, $\Delta\phi({\slashed{\vec{E}}_T}, \vec{p}_T^{j1,j2})$ and $S_T$ do not have significant impact on the results. 
For 3 b-jet final state, the most important kinematic variables are $\slashed{E}_T$, $H^b_T$, $m_{\rm inv}^{bb}$,  $\Delta R(b1, b2)$ and $S_T$, 
whereas $p_T^{b1}$, $p_T^{b3}$ and $\Delta\phi(\vec{p}_T^{bb}, \vec{p}_T^{b3})$ are also essential to obtain the best possible sensitivity. 
The other kinematic variables, namely $N_j$ and $\Delta\phi({\slashed{\vec{E}}_T}, \vec{p}_T^{b1,b2,b3})$, do not have significant impact on
the results. 

In order to showcase the sensitivity of our multivariate analysis, we choose some sample benchmark points with varied $m_H$ and $m_{\chi}$.
The $H$ is assumed to decay with 50\% branching ratio into each of the two available decay modes $\chi\chi$ and $b\bar b$. In order to assess the 
most optimistic scenario, the value of the Wilson coefficient is kept fixed at its perturbative limit, $\sqrt{4\pi}\sim$ 3.5 and the $H$-$b$-$\bar b$ 
coupling is computed such that $H$ can decay with the above mentioned branching ratios. The area under the Receiver Operating Characteristic (ROC) 
curve (AUC) is a good estimator of the efficiency of the BDT classifier in identifying the signal and background events. The AUC numbers corresponding 
to all the benchmark points along with the required integrated luminosity to obtain a $3\sigma$ statistical significance (and the 
corresponding $5\sigma$ numbers within parentheses) are quoted in  
Table~\ref{tab:res_mult} for the energy scale $\Lambda = 3$ TeV. The signal significance is 
calculated by ${\mathcal S}=\frac{S}{\sqrt{S+B}}$, where, $S$ and $B$ represents the number of signal and background events respectively.
\begin{table}[h]
\resizebox{\columnwidth}{!}{
\begin{tabular}{||c||c|c||c|c||}
\hline
\multicolumn{1}{||c||}{Benchmark} &
	\multicolumn{2}{|c||}{SR1(1 b-jet + $\slashed{E}_T$)} &
	\multicolumn{2}{|c||}{SR2(3 b-jet) } \\
\cline{2-5}
Points &  & Required &  & Required  \\
 & AUC & Luminosity (fb$^{-1}$) & AUC & Luminosity (fb$^{-1}$)  \\
\hline\hline
BP1         &0.91      &100.6      &0.94       &67.0                  \\
	($m_H = 600$ GeV, $m_{\chi} = 100$ GeV) &      &(279.4)      &       &(186.0)  \\
\hline
BP2         &0.94      &950.9      &0.88       &1836.7                 \\
	($m_H = 1400$ GeV, $m_{\chi} = 100$ GeV) &      &(2641.3)      &       &(5102.0)  \\
\hline
BP3         &0.94      &578.7      &0.92       &656.3                  \\
	($m_H = 1200$ GeV, $m_{\chi} =500$ GeV) &      &(1607.6)      &       &(1823.0)  \\
\hline
BP4         &0.95      &2123.3     &0.86       &7694.7                 \\
	($m_H = 1800$ GeV, $m_{\chi} = 500$ GeV) &      &(5898.1)      &       &($>$ 10000)  \\
\hline\hline
\end{tabular}
}
\caption{Results of our BDT analysis performed using the XGBoost toolkit corresponding to four sample benchmark points 
for both the signal regions. For all these benchmark points, 
$\Lambda = 3$ TeV and BR($H\to\chi\chi$) = BR($H\to b\bar b$) = 50\%. The Wilson coefficient $f$ is 
kept fixed at its largest perturbative limit $\sqrt{4\pi}\sim$ 3.5. The luminosity column shows the required luminosity 
to obtain $3\sigma$ ($5\sigma$) statistical significance.}
\label{tab:res_mult}
\end{table}

Evidently, with $\Lambda=3$ TeV, one can probe a large portion of the parameter space. One can easily have a $3\sigma$ evidence for $m_H$ up to 2 TeV,
and discovery $\sim$ 1.5 TeV for both $m_{\chi} =100$ GeV and 500 GeV through SR1. SR2 performs at par with SR1 or even better when $m_H$ is on the lighter side. 
As the mass gap between $m_H$ and $m_{\chi}$ increases, SR1 performs better. This is mostly due to the presence of a large 
$\slashed{E}_T$ in SR1. Clearly, a definite discovery significance of $5\sigma$ can be attained for BP1, BP2 and BP3 through SR1 and 
BP1 and BP3 through SR2. We observed that in order to produce BR($H\to\chi\chi$) = BR($H\to b\bar b$) = 50\% with a 
$H$-$\chi$-$\chi$ coupling $\sim 1$, one requires the $H$-$b$-$\bar b$ coupling to be $\sim 10^{-3}$-$10^{-4} \times y_{b\bar bh}|_{SM}$ 
for the benchmark points. This renders the production cross-section $\sigma (pp\to bH)$ solely reliant on the dimension-6 operator, since the 
contribution from the s-channel mediated diagram in Fig.~\ref{fig:proc} is negligibly small in comparison. 

\begin{figure}[h!]
\includegraphics[width=7.0cm,height=6cm]{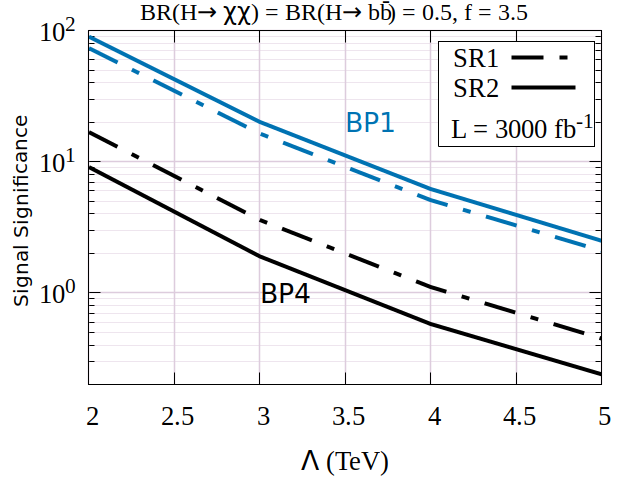}
\caption{The HL-LHC reach in signal significance in the two signal regions, SR1 (dashed line) and SR2 (solid line) as a function 
of $\Lambda$ for BP1 (blue line) and BP4 (black line). The other two benchmark points lie in between.} 
\label{fig:hllhc_sign}
\end{figure}

The cross-section drops rapidly with increasing $\Lambda$ and drops by one order of magnitude when increased from 3 TeV to 5 TeV.
In Fig.~\ref{fig:hllhc_sign} we show how much the sensitivity can change with the variation of the new physics scale.
The y-axis shows the expected signal significance in both SR1 and SR2 for the highest possible luminosity of $3000~{\rm fb}^{-1}$ 
at 14 TeV LHC. The x-axis shows variation of $\Lambda$. We show the results for BP1 and BP4 since these are the lightest and 
heaviest benchmark points respectively. Clearly, for BP1 at $\Lambda=5$ TeV, one can barely obtain a $3\sigma$ statistical significance 
in SR2 and $2\sigma$ in SR1. The trend reverses as the benchmark points become heavier. Now SR1 is the more 
sensitive one. However, for $\Lambda=5$ TeV, BP4 as well as BP2 and BP3 has poor sensitivity in either of the signal regions. We 
checked that even BP2 can barely achieve a signal significance of $1\sigma$ in that case. 
\subsubsection*{Comparison to cut-based analysis}
In comparison to the multivariate analysis, a canonical cut-based analysis performs poorly. To illustrate this fact, we implemented a set of cuts 
carefully designed to maximize the efficiency in both the signal regions. For SR1, we demand that the final state must have only one $b$-jet with $p_T^b > 250$ GeV, at least two light jets with $p_T^j < 50$ GeV, no 
charged leptons, ${\slashed E}_T > 250$ GeV, $\Delta\phi({\slashed{\vec{E}}_T}, \vec{p}_T^b) > 1.0$, $M_{EFF} > 500$ GeV and $S_T < 0.1$. For SR2, the final state must have at least three $b$-jets with the 
hardest two having $p_T^b > 150$ GeV, no charged leptons, no light jets with $p_T^j > 50$ GeV, 
${\slashed E}_T < 100$ GeV, $|m_{\rm inv}^{bb} - 125.0| > 25$ GeV, $H_T^b > 500$ GeV, $\Delta\phi(\vec{p}_T^{bb}, \vec{p}_T^{b3}) > 1.0$ and $S_T > 0.1$. With these set of cuts, we inspect the four benchmark
points and the aforementioned background channels. The results are presented in Table~\ref{tab:res_cut_sig}. 
We observed that through such a cut-based analysis, for BP1
one can attain a $3\sigma$ ($5\sigma$) signal significance at an integrated luminosity of 214.2 
(595.1) ${\rm fb}^{-1}$ and 500.1 (1389.2) ${\rm fb}^{-1}$ in SR1 and SR2, respectively. BP3 can only be 
probed in SR1 close to a signal significance of $3\sigma$. BP3 cannot be probed in SR2 within the luminosity reach. BP2 and BP4 remain 
completely out of reach in both the signal regions. 
Thus, it is evident that the multivariate BDT analysis offers a distinct advantage over the cut-based analysis. Note that, for 
the chosen set of optimized cuts SR1 always has better sensitivity compared to SR2. This is in contrast to our multivariate analysis results
which clearly show that the relative sensitivity is reversed for BP1. It turns out that SR2 is more effective within a small parameter space
highlighting the fact that the BDT algorithm is more sensitive to the subtle changes in kinematics in order to identify the signal events.
This feature is discussed in more detail in the subsequent section.  
\begin{table}[h]
\scriptsize
\begin{tabular}{||c||c|c||}
\hline
\multicolumn{1}{||c||}{Benchmark} &
\multicolumn{2}{|c||}{Required Luminosity (fb$^{-1}$) } \\
\cline{2-3}
Points & SR1(1 b-jet + $\slashed{E}_T$)  & SR2(3 b-jet)    \\  
\hline\hline
BP1          &214.2 &500.1                  \\
\hline
BP2          &9305.1 & $>$10000                  \\
\hline
BP3          &3748.4 & $>$10000                \\
\hline
BP4          & $>$10000 & $>$10000                 \\
\hline\hline

\end{tabular}
\caption{Results of cut-based analysis for the four sample benchmark points for both the signal regions.  
For all these benchmark points, $\Lambda = 3$ TeV and BR($H\to\chi\chi$) = BR($H\to b\bar b$) = 50\%. The Wilson coefficient $f$ is 
kept fixed at its largest perturbative limit $\sqrt{4\pi}\sim$ 3.5. The luminosity column shows the required luminosity 
to obtain $3\sigma$ statistical significance.}
\label{tab:res_cut_sig}
\end{table}

\subsection{Discovery reach at HL-LHC}
\label{sec:disc}
\begin{figure}[h!]
\includegraphics[width=7.0cm,height=6cm]{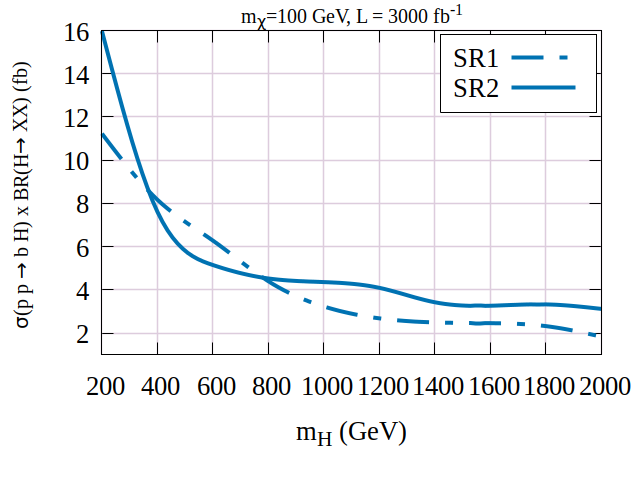}
\includegraphics[width=7.0cm,height=6cm]{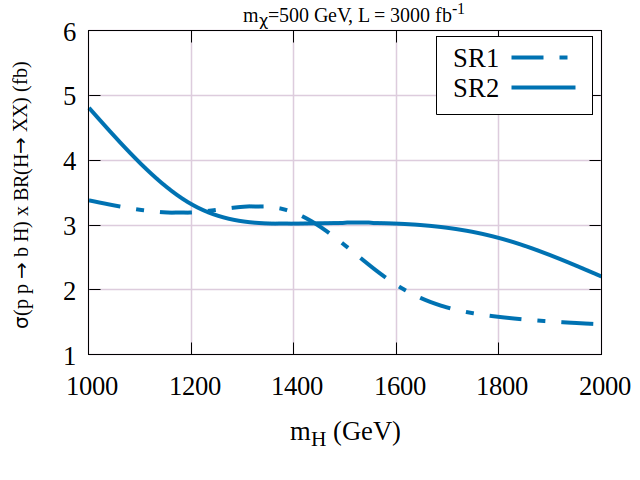}
\caption{The HL-LHC reach in $\sigma (pp\to b H)\times {\rm BR}(H\to XX)$, where $XX\equiv\chi\chi$ for SR1 
(dashed line) and $b\bar b$ for SR2 (solid line) as a function of the heavy scalar mass, $m_H$, in order to achieve a 
$3\sigma$ statistical significance. The results are obtained through our BDT analyses performed using the XGBoost toolkit.}
\label{fig:disc_sigbr}
\end{figure}

We now proceed to assess the extent to which the 14 TeV LHC can probe the present scenario at the highest possible projected luminosity, 
$3000~{\rm fb}^{-1}$. For this, we study two sample values of $m_{\chi}=100$ and 500 GeV. 
Fig.~\ref{fig:disc_sigbr} shows the required $\sigma (pp\to bH )\times {\rm BR}(H\to\chi\chi)$ and 
$\sigma (pp\to b H)\times {\rm BR}(H\to b\bar b)$ as a function of $m_H$ for SR1 and SR2, respectively,
such that a $3\sigma$ statistical significance is obtained at $3000~{\rm fb}^{-1}$.  
The $m_H$ threshold is chosen such that $H\to\chi\chi$ decay mode is open. Understandably, one can 
obtain non-zero event rate for SR2 even below this $m_H$ threshold. One can simply look at the existing experimental studies looking 
for non-standard Higgs bosons decaying into a pair of $b$-quarks for such limits, see e.g. \cite{Khachatryan:2015tra}. 
For both $m_{\chi}=100$ GeV and 500 GeV, the trend looks same. At the low $m_H$ region as well as high $m_H$ region, SR1 has better 
sensitivity compared to SR2 which is only more efficient within an intermediate range. In the low $m_H$ region, 
kinematics of the final state $b$-quarks are not distinctly different from their SM counterpart in SR2. Expectedly, the situation improves 
dramatically with increasing $m_H$ before the sensitivity saturates. On the other hand, SR1 is always associated with a large ${\slashed E}_T$ 
which helps it gain better sensitivity specially in the low $m_H$ region. ${\slashed E}_T$ along with the angular variables only get more and more 
distinctive with increasing $m_H$, making SR1 the more sensitive channel at heavier mass region as well. The sensitivity in the 1-2 TeV range of $m_H$ 
is very similar for both the two different choices of $m_{\chi}$ indicating the fact that mass of the invisible particle does not have 
a severe impact on the outcome. It should be noted that the results in Fig.~\ref{fig:disc_sigbr} depend on $m_H$ only. Whatever choice of $\Lambda$ predicts a particular $\sigma (pp \to bH) \times {\rm BR}(H \to XX)$ for each $m_H$ is similarly favourable, so long as the Lorentz structure of the effective interaction is the same, since the latter has a role in deciding the efficiency of the BDT analysis.

It is worth mentioning that the experimental collaborations are yet to probe the signal region SR1 in the context of 
heavy scalar search. However,
they have studied the multi $b$-jet final state SR2 \cite{Sirunyan:2018taj,Aad:2019zwb} as mentioned before. These studies combine 
the four and five flavor contributions to the cross-section of a signal consisting of at least three $b$-jets and present their 
exclusion limit on $\sigma(pp\to b\bar bH)\times ~{\rm BR}(H\to b\bar b)$ 
as a function of $m_H$ with $\sqrt{s}=13$ TeV for an integrated luminosity $35.7~{\rm fb}^{-1}$ \cite{Sirunyan:2018taj} and 
$27.8~{\rm fb}^{-1}$ \cite{Aad:2019zwb}. In the context of this work, $\sigma(pp\to b\bar bH)\times ~{\rm BR}(H\to b\bar b)$ 
is negligibly small by virtue of a small $Hb\bar b$ coupling as discussed in section~\ref{sec:obj}. However, one would expect a similar 
exclusion limit for $\sigma(pp\to bH)\times ~{\rm BR}(H\to b\bar b)$ as well. Even if we assume in a conservative way, that 
both these exclusion limits lie in the same ballpark, our benchmark points still cannot be excluded with the present experimental 
sensitivity. Moreover, we present in Fig.~\ref{fig:disc_sigbr} the highest possible sensitivity that can be reached at the LHC 
in order to attain a $3\sigma$ statistical significance. The required 
$\sigma(pp\to bH)\times ~{\rm BR}(H\to b\bar b)$ in this figure over a wide range of $m_H$ lies at least one order of magnitude below 
the present experimental sensitivity presented in \cite{Sirunyan:2018taj,Aad:2019zwb}.

\begin{figure}[h!]
\includegraphics[width=7.0cm,height=6cm]{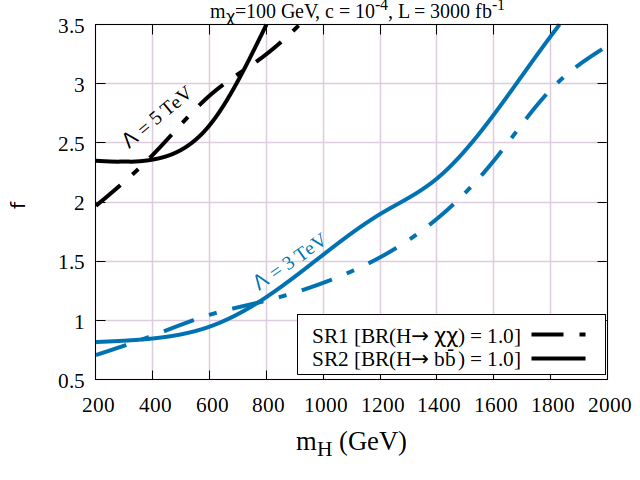}
\caption{The distribution of the Wilson coefficient, $f$ as a function of $m_H$ for SR1 (dashed line) and SR2 (solid line) 
in order to produce the required $\sigma (pp\to b H)\times {\rm BR}(H\to XX)$ resulting in a 
$3\sigma$ statistical significance at HL-LHC for $m_{\chi}=100$ GeV, $c= 10^{-4}$. BR($H\to\chi\chi$) = BR$(H\to b\bar b)$ = 1.0 
respectively for SR1 and SR2. The results are shown for $\Lambda=3$ TeV (blue line) and 5 TeV (black line).}
\label{fig:disc_wilf}
\end{figure}

We use the results presented in Fig.~\ref{fig:disc_sigbr} to ascertain the required values of the Wilson coefficient, 
$f$, that can produce the projected $\sigma (pp\to b H)\times {\rm BR}(H\to XX)$. While doing so, we take note of the fact that there are 
two competing production channels. As a result, the choice of $Hb\bar b$ vertex factor plays a crucial role. Since our goal is to 
determine the accessible range of $f$, we choose a parameter space such that the $Hb\bar bg$ effective vertex diagram contributes 
dominantly to the cross-section. Therefore, we keep the $Hb\bar b$ vertex factor, $V_{Hb\bar b} = c\times y_{b\bar bh}|_{SM}$ small. 
Here $h$ is the 125 GeV Higgs boson and $c$ is just a scale factor. We keep $c=10^{-4}$ to keep this coupling sufficiently small.  
In order to ascertain the maximum impact of our analysis, we keep BR($H\to\chi\chi$) = 1.0 and BR$(H\to b\bar b)$ = 1.0  
for SR1 and SR2 respectively. The results are shown in Fig.~\ref{fig:disc_wilf} for $m_{\chi}=100$ GeV and $\Lambda=3$ and 5 TeV.
The upper limit of $f$ is restricted to the perturbative limit. The coupling strength of $H$ to the pair of invisible particles
$V_{H\chi\chi}$ is fixed within its perturbative limit such that the criteria on the two relevant branching ratios are met. 
Clearly, SR1 has a greater sensitivity compared to SR2 apart from a small part of the parameter region as also seen in 
Fig.~\ref{fig:disc_sigbr}. For $\Lambda=3$ TeV, heavy higgs mass of $\sim$ 2 TeV can be easily probed at the HL-LHC up to a 
signal significance of $3\sigma$. The Wilson coefficient, $f$ can be probed all the way down to $\sim 0.6$. The sensitivity 
worsens understandably with increasing $\Lambda$ and as evident, one can only probe $m_H \lesssim 1$ TeV and $f\gtrsim 1.9$.   
Note that, for a large enough $c \gtrsim 10^{-2}$, the production cross-section will be dominated by the s-channel $b$-quark 
mediated diagram. In that case, SR2 will be the favoured signal region to probe the scenario throughout the parameter space. 
However, that does not improve the sensitivity in probing smaller values of $f$ unless the $Hb\bar b$ coupling is measured to a 
high degree of precision such that one can discern the comparatively much smaller contribution to the event rate arising from 
the dimension six operator. Here we discussed the $m_{\chi}=100$ GeV case as an example to showcase the extent of the 
sensitivity of our multivariate analysis. The accessible range of $f$ can be quite different for other choices of $m_{\chi}$. 

\section{Summary and conclusion}
\label{sec:concl}
In this work we have explored two signal regions through multivariate analysis to probe the prospect of a heavy scalar 
being singly produced in association with a $b$-quark at the 14 TeV LHC. The production channel $bg\to bH$ can get contribution from 
an effective interaction of the form $Hb\bar bg$. Such an interaction can be generated from viable dimension six operators. However, 
one needs to treat the $b$-quarks as partons in order for this interaction to contribute to the cross-section starting from a proton-proton 
collision. Instead of exploring some specific new physics model we parameterize its effect through the interaction strength and ascertain 
the sensitivity of the LHC to probe this scenario at its peak projected luminosity. In order to do so, we identify the most probable visible
decay mode of $H$ along with an invisible decay mode to construct two signal regions. Looking at the kinematics of the final states we 
choose two sets of kinematic variables to carry out our multivariate analysis using the XGBoost toolkit. We show that even though the 
signal regions have potentially large SM backgrounds, one can still achieve good sensitivity through this kind of multivariate analysis.  
We determine the required signal cross-sections in order to obtain $3\sigma$ statistical significance at the 14 TeV LHC with an integrated 
luminosity of $3000~{\rm fb}^{-1}$. We present the result as function of $m_H$, which is model independent and is applicable to any new 
physics model giving rise to similar interactions. We further use this information to determine the efficiency of the LHC in probing the 
Wilson coefficient $f$ of the effective interaction. We find that $m_H\lesssim 2$ TeV can easily be probed with a $3\sigma$ statistical 
significance staying within the perturbative 
limit of $f$ with the new physics scale set at 3 TeV. The situation worsens considerably with $\Lambda = 5$ TeV, the reach in $m_H$ being 
restricted to below the TeV range. The probe on $f$ is dependent on the choice of $Hb\bar b$ interaction strength. Our focus being the 
new effective interaction, we keep the $Hb\bar b$ coupling strength to be small.  
We observe that $f < 0.5$ is difficult to probe in either of the two signal regions. 
The results also show that in presence of such an effective interaction one can in principle probe much smaller $Hb\bar b$ coupling 
compared to the existing experimental studies. 
\section{Acknowledgements}
BM acknowledges the hospitality of Katri Huitu and the Helsinki institute of Physics during the early part of the project. 
The research of BM till November, 2019, was supported by the Department of Atomic Energy, Government of India, through
funding made available to the Regional Centre for Accelerator-based Particle Physics, Harish-Chandra Research Institute, Allahabad.

\end{document}